\documentclass{aa}  

\usepackage{graphicx}
\usepackage{txfonts}
\usepackage[colorlinks=true,
            linkcolor=blue,
            urlcolor=blue,
            citecolor=blue]{hyperref}
\newcommand{\hmsun}{h^{-1}{\rm M}_\odot}
\newcommand{\hmpc}{h^{-1}{\rm Mpc}}

\begin{document} 
  
   \title{How do galaxies populate halos in extreme density environments?}
   \subtitle{An analysis of the Halo Occupation Distribution in SDSS}

   \author{Ignacio G. Alfaro \thanks{E-mail:german.alfaro@unc.edu.ar}, Facundo Rodriguez, Andr\'es N. Ruiz, Heliana E. Luparello \& Diego Garcia Lambas }
   
   \authorrunning{I. G. Alfaro et al.}
   
   \institute{Instituto de Astronomía Teórica y Experimental, CONICET-UNC, Laprida 854, X5000BGR, C\'ordoba, Argentina \\ Observatorio Astron\'omico de C\'ordoba, UNC, Laprida 854, X5000BGR, C\'ordoba, Argentina.}

   \date{\today}

  \abstract
   {Recent works have shown that the properties of galaxy populations in simulated  dark matter halos vary with large-scale environments. These results suggest a variation in the halo occupation distribution (HOD) in extreme density environments since the dynamical and astrophysical conditions prevailing in these regions may significantly affect the formation and evolution of their halos and residing galaxies, influencing the mean number of satellite galaxies. To analyse these effects, we identify cosmic voids and future virialised structures (FVS) in the Sloan Digital Sky Server Data Release 12 (SDSS-DR12) and estimate the HOD within these super-structures using group catalogues as dark matter halo proxies.}
   {Our goal is to use observational galaxy data to characterise the HOD within cosmic voids and FVS,  explore the different properties of these galaxies populations and compare them with the general results outside these super-structures.}
   {We use a publicly available observational galaxy catalogue with information on redshifts, positions, magnitudes and other astrophysical features to build a volume complete galaxy sample and identify cosmic voids and FVS. Using a publicly available galaxy group catalogue as a proxy to dark matter halos, we compute the HOD within both types of super-structures for different absolute magnitude thresholds. We also study the dependence on the results on the main void and FVS properties, density and volume. We also analysed the main characteristics of the stellar content of galaxies inside these extreme-density regions such as the mean stellar age and the galaxy light concentration index. In all cases, we compare the results with those derived from the Field sample, defined by objects outside both types of environments.}
   {Inside cosmic voids, we find a strong decrease in HOD concerning the Field results. In the most extreme cases, the mean number of satellites fall to $\sim 50\%$. Inside FVS, the HOD shows a significant increase to the Field, with a $\sim 40\%$ excess in the mean number of satellites. These results are present for the different galaxy luminosity ranges explored. In both environments, the differences with respect to the Field increases for the extreme values of the density environments. However, we obtain no signs of variations related to intrinsic characteristics of the super-structures, indicating that the effects depend mainly on the density of the large-scale environment. In addition, we find that the cumulative distribution of the mean age of stars of the central galaxy also varies in the different regions, this suggests that the history of the formation of the dark matter halos may be different. Finally, we explore the HOD for the 25\%  youngest (oldest) galaxies, based on the mean age of their stars. We find that for the low-mass groups the youngest galaxies are only present inside voids, and are generally central galaxies. On the other hand, for the high-mass groups the FVS environments show the same increase in the HOD concerning the Field as previously mentioned. We find that cosmic voids lack a significant fraction of galaxies with the oldest stellar population.}
  {}

   \maketitle
  
%

\section{Introduction}
\label{sec:introduction}

The current paradigm of structure formation in the Universe predicts that galaxies form within virialised dark matter halos as a product of accretion of baryonic material.
However, the variety of complex astrophysical phenomena involved in the galaxy formation and evolution process makes it difficult to determine unambiguously how galaxies populate a given halo.
Understanding this relationship is key to understanding the formation and evolution of large-scale structures, as well as its influence on the properties of galaxies.

A valuable statistical tool to study the connection between galaxies and their dark matter halos is the Halo Occupation Distribution (HOD). 
The HOD is defined as the probability distribution that a virialised halo of mass $M_{\rm halo}$ contains $N$ galaxies with specific characteristics, $P(N|M_{\rm halo})$. 
It is generally assumed that, at first order, the HOD depends only on the mass of the halo \citep[e.g.][]{Jing1998,Ma2000, Peacock2000,Seljak2000,Scoccimarro2001, Berlind2002,Cooray2002,Berlind2003, Zheng2005, Yang2007, Rodriguez2015, Rodriguez2020}.
However, recent works in simulations has shown a correlation between the HOD and the density of the environment in which the halos evolve \citep[e.g.][]{Zehavi2018, Artale2018, bose_hod_2019}.
This led to the study of the HOD behaviour within regions with extreme density values, such as cosmic voids \citep{Alfaro2020} and Future Virialised Structures  \citep[FVS,][]{Alfaro2021}.

The large-scale structure of the Universe, usually called the cosmic web, is the result of mass accretion, a process dominated mainly by gravity. 
As is well known, this gives rise to regions where the density of matter reaches extreme values concerning the average density. 
The cosmic voids correspond to the regions with the lowest density, while the FVS correspond to the highest. 
Although there are many definitions for a void, most agree that these regions comprise most of the volume of the Universe, contain a small fraction of galaxies, which, added to their expanding dynamics, make gravitational interactions between objects infrequent within them, affecting the growth and development of the structure \citep{ceccarelli_voids_2006,Patiri2006,Colberg2008, Pan2012, Hoyle2012, ruiz_void_2015, ruiz_into_2019}. 
On the other hand, it is well known that mass flows from the less dense regions to the denser ones, mainly through filaments and walls. 
At the intersections of these two structures, nodes can form, which under certain conditions can evolve into the densest virialised regions of the Universe, called FVS.
Observationally, the properties of galaxies and groups in these extremely dense environments, which also contain most of the high-mass halos, suggest that galaxy groups may have formed earlier in these super-structures than in the middle regions of the Universe \citep{einasto_1997,einasto_2001,Dunner:2006,Einasto:2007,costa_duarte,luparello_fvs_2011,liivamagi}. 

In \cite{Alfaro2020} and \cite{Alfaro2021} we found evidence of significant variations in HOD within voids and FVS, respectively. 
For this, we use both semi-analytical and hydrodynamical simulations. 
We find that there is a correlation between the age of halo formation, the average number of galaxies in a halo, and the environment in which they are located. 
The halos within the voids had a lower than average HOD and formed at lower than average redshift.
Whereas, within the FVS the halos had a higher than average HOD and formed at higher redshift. 
This is indicative, as observed in the synthetic data, that the halos in these regions have evolutionary histories different from the average, which affects how galaxies populate them. 
The methods with which we identify voids and FVS are fully reproducible observationally.

Taking advantage of the large data volume provided by the Sloan Digital Sky Survey Data Release 12  \citep[SDSS-DR12,][]{Alam2015}, in this work, we set out to explore the HOD in extreme density environments to assess whether the results obtained theoretically correspond to those from observations. 
To meet this objective, we also use the SDSS-DR12 group catalogue developed by \cite{Rodriguez2020} and also the possibility to identify extreme environments of the large-scale structure through our algorithms.
Voids were identified with the algorithm of \cite{ruiz_void_2015}, while FVS were detected following \cite{luparello_fvs_2011}.

This paper is organised as follows.
In Sec. \ref{sec:data}, we describe the observational galaxy catalogue and the sample of objects used in this work. We also characterized the galaxy groups catalogue and the algorithms to identify cosmic voids and FVS.
In Sec. \ref{sec:est_prop}, we show the main properties of our super-structures catalogues. %
In Sec. \ref{sec:results}, we describe the method we used to estimate the HOD and define the three samples of galaxy groups that we analysed: groups in Voids, FVS and Field. We present and compare the results of the HOD measurements for these three different regions. In this section, we also explore the dependence of the results with the density of the environment surrounding the group and with the intrinsic properties of the super-structures.
In Sec. \ref{sec:stellar}, we compare the onset time of star formation of the central galaxies inside the voids and FVS with the Field results. Based on this measure of time, we also computed the HOD for the $\sim 25\%$ of the youngest and oldest galaxies.
Finally, in Sec. \ref{sec:conclusions}, we present our summary and conclusions.


\section{Data}
\label{sec:data}

In this section, we describe the galaxy catalogue, the galaxy group finder with their dark matter halo mass estimation and the voids and FVS identification algorithms used in this work.
\subsection{The SDSS galaxy catalogue}
\label{sec:sdss}

We use the main galaxy sample of the Sloan Digital Sky Survey Data Release 12 \citep[SDSS-DR12;][]{Alam2015}. 
This Legacy footprint area covers more than 8400 deg$^2$ in five optical bandpasses and has more than $\sim$800 000 million galaxies with redshift up to $z=0.3$ and apparent magnitudes in the $r-$band lower than 17.77.
In addition to the redshift, position and magnitudes, we employed the astrophysical data from the Portsmounth method with star formation model to the galaxies and stellar masses estimated following \cite{Maraston2006} method. 
This estimation fits stellar evolution models to the SDSS photometry, using BOSS redshifts.
The star-formation model considers metallicity and three star-formation histories: constant, truncated, and exponentially declining ($\tau$), which is provided in the 'SFH' column.
The 'age´ parameter listed gives the initial time for the onset of star-formation in each model.
In this table, we assume the Kroupa IMF.\footnote{This information was extracted from \url{skyserver.sdss.org/dr12/}, further details of this data can be found in \textit{stellarMassStarformingPort} table. }
From this sample we select galaxies with a limiting redshift, $z_{\rm lim} = 0.1$ and a limiting $r$-band absolute magnitude of $M_{\rm r} - 5\log_{\rm 10}(h) = -19.77$.
Thus, our sample is complete in volume providing accurate tracer galaxies to identify suitable super structures, ie. voids and FVS. 

\subsection{The group galaxy catalogue}
\label{sec:groups}

To compute the HOD, in addition to the photometric data of the galaxies, we need to associate galaxies with the dark matter halos they inhabit and determine the masses of these halos.
For this purpose, we use the galaxy groups catalogue presented by \cite{Rodriguez2020}.
This group sample was obtained by a new iterative algorithm that combines the friends-of-friends \citep{Merchan2005} and halo-based \citep{Yang2007} techniques. 
If the group members vary, the method recalculates the dark matter halo properties, repeating the process until no more changes in the groups are needed. 
This approach allows maintaining a high performance both to detect low and high numbers of members systems. 
As part of the process, this procedure provides a halo mass estimation for each group, $M_{\rm group}$, which is obtained by an abundance matching technique based on luminosity, i.e. assuming a one-to-one relationship between the characteristic luminosity of the group and the halo mass \citep{Vale2004, Kravtsov2004, Conroy2006,behroozi_2010}. 

Among other advantages, it was found that this galaxy group sample presents an excellent agreement between the mass it provides and those obtained by weak gravitational lensing techniques \citep{Gonzalez2021}.
In addition, it showed a high efficiency at comparing properties of central and satellite galaxies with results obtained in simulations \citep{Rodriguez2021}.

Our final galaxy catalogue is a volume-limited sample of galaxies up to $z=0.1$ comprising 134405 objects with angular positions, spectroscopic magnitudes and other astrophysical data, together with their corresponding group and halo host membership and mass.

\subsection{Voids identification}
\label{sec:voids_id}

We use the algorithm presented in \citet{ruiz_void_2015} to identify spherical cosmic voids in the SDSS-DR12 main galaxy sample described in Sec. \ref{sec:sdss}.

Using galaxies as density tracers, we measure the integrated density contrast profile ($\Delta$) in all underdense regions and we identify the largest sphere satisfying $\Delta(R_{\rm void}) < \Delta_{\rm lim}$, with $R_{\rm void}$ the void radius, and $\Delta_{\rm lim}$ an integrated density contrast threshold set to $-0.9$, i.e., our voids contains 10 percent of the mean density of tracers.
All spheres that satisfy this condition are then cleaned up by removing superposition and prioritizing the largest sphere.

It is worth to mention that, in order to take into account survey boundaries and holes present in SDSS-DR12 , the void identification process considers an angular mask of the observational data constructed using \textsc{HEALPix} \citep{HEALPIX}.
Also, none of the voids identified in the boundaries of the catalogue was considered. 

\subsection{FVS identification}
\label{sec:fvs_id}

As stated by the $\Lambda$CDM Concordance Cosmological Model, the accelerated expansion dominates the present and future dynamics of the Universe.
Within this framework, the FVS are defined as the largest overdense systems that will remain bound and go through its viralization process during its subsequent evolution.
Thus, the identification of FVS is based on a procedure that search for current overdense regions that also must satisfy the condition of evolving as connected systems.
The details are given in \citet{luparello_fvs_2011}, who combine the observational method of the luminosity density field \citep{Einasto:2007} with the theoretical criteria of the mass overdensity for a structure to remain bound \citep{Dunner:2006}.
The main advantage of this procedure is that it can be easily applied in both observational and numerically simulated galaxy data.
In order to identify FVS, we first construct a luminosity density field by convolving the spatial distribution of the galaxies with a kernel function weighted by galaxy luminosity.
This procedure provides a continuous luminosity density map across the analysed volume, with a resolution  set by cubic cells of $1~\hmpc$ side.
Then, we applied a percolation algorithm that allow us to select the connected cells above certain luminosity threshold.
In order to be considered part of a structure, each cell must satisfy $\delta L_{\rm loc}=\rho_{\rm lum}/\bar{\rho}_{\rm lum}\geq 5.5$, where $\rho_{\rm lum}$ is the luminosity density of the cell, and $\bar{\rho}_{\rm lum}$ is the mean luminosity density of the set of cells.
As result of this procedure we obtain the list of the cells belonging to each FVS,
which allow us to directly identify its galaxy members.
We also imposed $10^{12} h^2 L_{\odot}$ as a lower limit for the total FVS luminosities, 
avoiding contamination from smaller systems. 

\section{Properties of super-structure catalogues}
\label{sec:est_prop}

In this section, we give a brief description of the main properties of the observational FVS and voids catalogues that we use throughout the work.
We present both the characteristics of the super-structures as a whole and those of the galaxy groups that compose them.

\begin{figure}[h!]
\begin{center}
\includegraphics[width=\columnwidth]{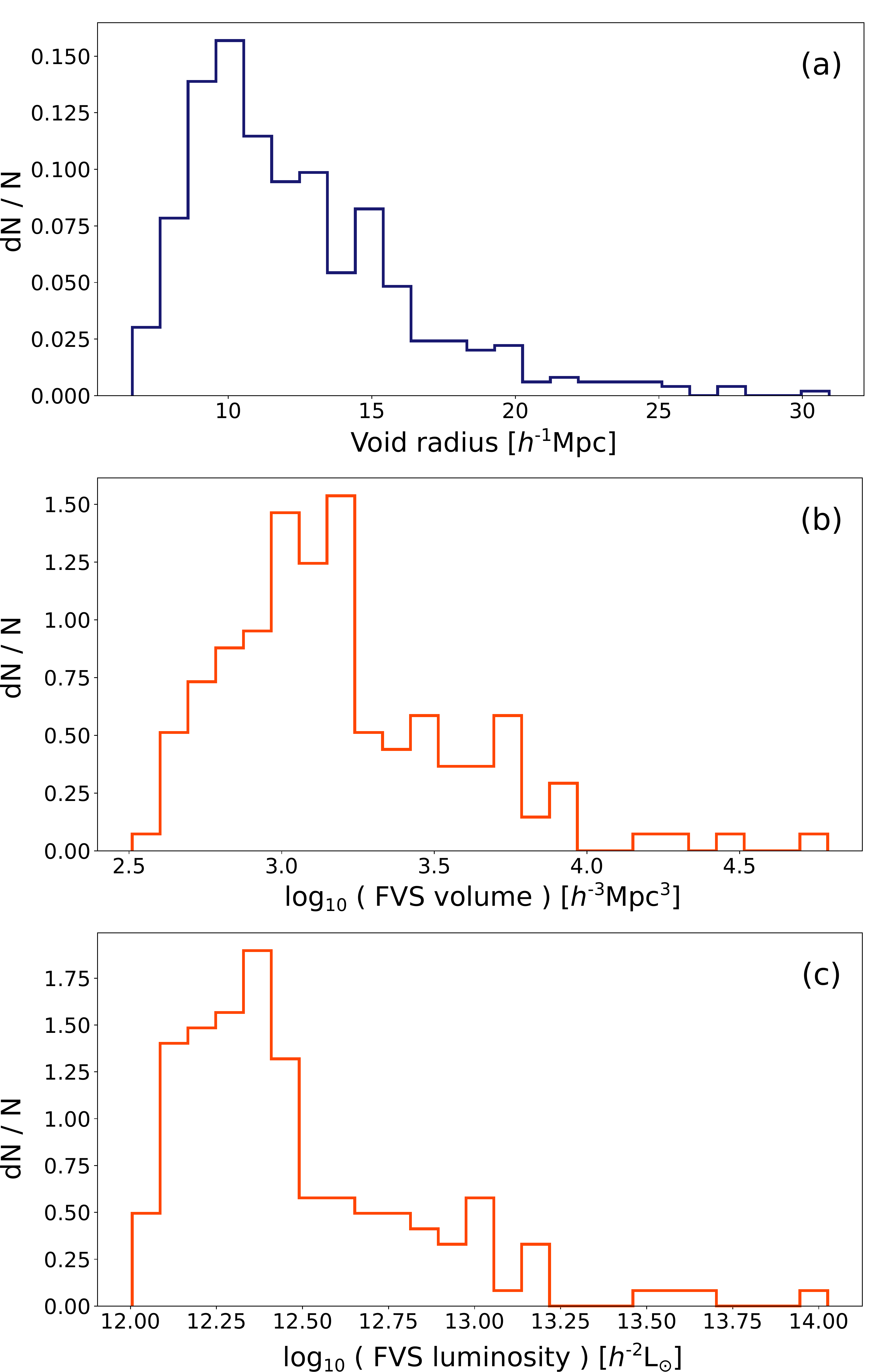}
\end{center}
\caption{\label{fig:str_prop} Normalized distributions of properties of the super-structures identified in the SDSS-DR12. Void radii are in panel (a), FVS volumes in panel (b) and FVS luminosities in panel (c).}
\end{figure}

We apply the super-structure identification algorithms described in Sec. \ref{sec:voids_id} for voids and \ref{sec:fvs_id} for FVS to the full galaxies sample presented in Sec. \ref{sec:sdss}.
We find 512 voids and 150 FVS, which contain 2041 and 18355 galaxies respectively.
In both cases, the identified regions show a wide variety of volumes. 
This is evident in the distribution of void sizes, shown in panel (a) of Fig. \ref{fig:str_prop}, and the distribution of FVS volumes shown in panel (b) of the same figure.
While both structures show broad volume distributions, the FVS span ranges of several orders of magnitude.
For the FVS we further calculate the total luminosity, whose distribution is shown in panel (c) of Fig. \ref{fig:str_prop}, where it can be seen that they also cover a wide range of values.
%

\begin{figure}[h!]
\begin{center}
\includegraphics[width=\columnwidth]{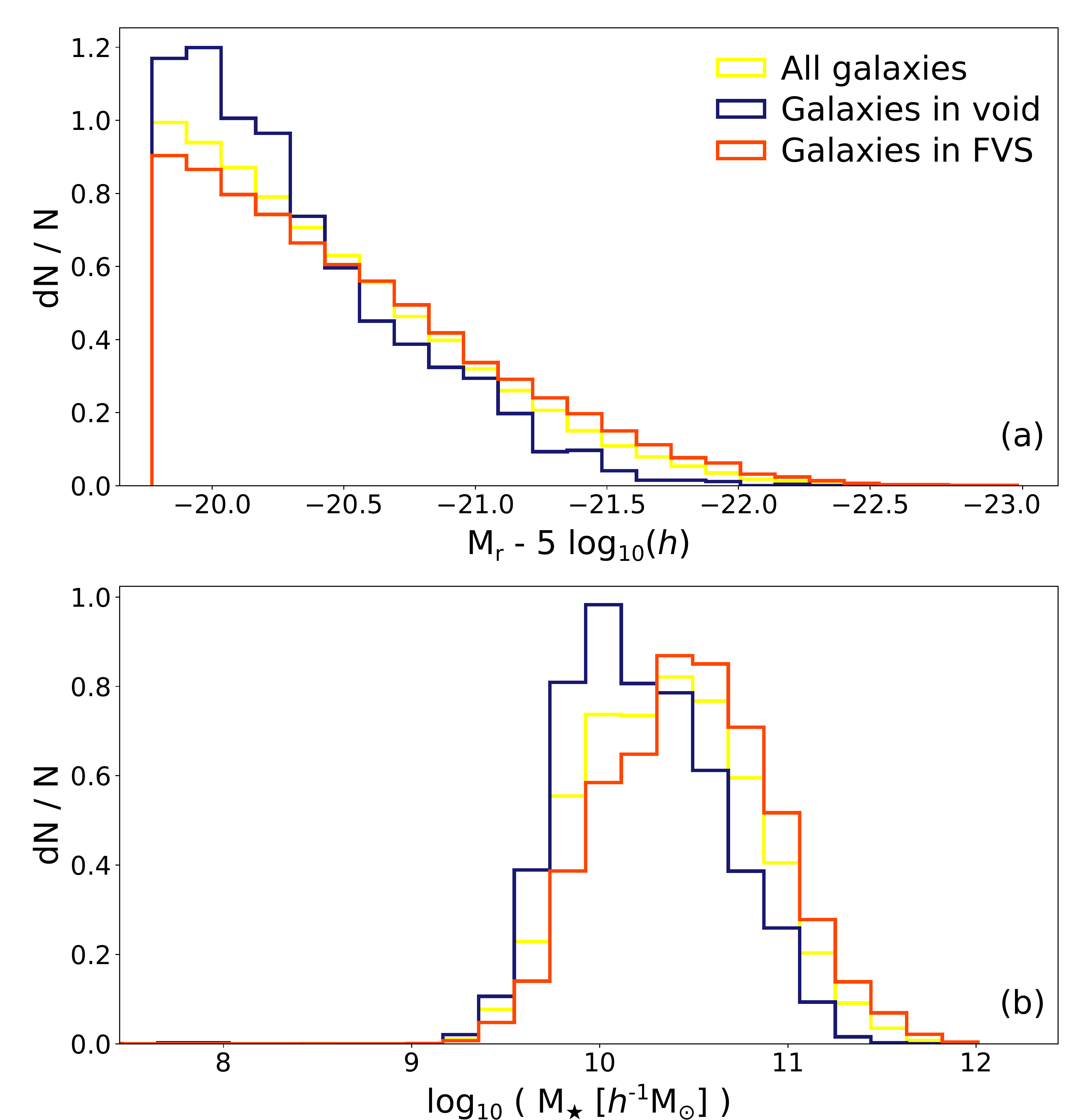}
\end{center}
\caption{\label{fig:hist_gal} Normalized distributions for properties of galaxies of the SDSS-DR12 used in this work. In panel (a) we show the $r$-band absolute magnitude and in panel (b) the stellar mass. In both cases yellow lines corresponds to all galaxies in the catalogue, blue lines to galaxies inside cosmic voids and red lines to galaxies residing in FVS, as indicated in the key.}
\end{figure}

Regarding the general properties of the galaxies populating these regions, Fig. \ref{fig:hist_gal} shows in panel (a) the distributions of the $r$-band absolute magnitudes $M_{\rm r} - 5\log_{\rm 10}(h)$ for the complete sample of galaxies (in yellow), galaxies in voids (in blue) and in FVS (in red).
As expected, the voids galaxies show an excess of faint galaxies (with $M_{\rm r} - 5\log_{\rm 10}(h) > -20.5$) concerning the full galaxy sample.
On the other hand, the FVS have a higher proportion of bright galaxies ($M_{\rm r} - 5\log_{\rm 10}(h) < -21$) than the mean.
These differences in the galaxy populations can also be reflected in the stellar mass distribution, as shown in panel (b) of Fig. \ref{fig:hist_gal}, which has the same colour pattern as the above panel to distinguish the object samples.

\begin{figure}[h!]
\begin{center}
\includegraphics[width=\columnwidth]{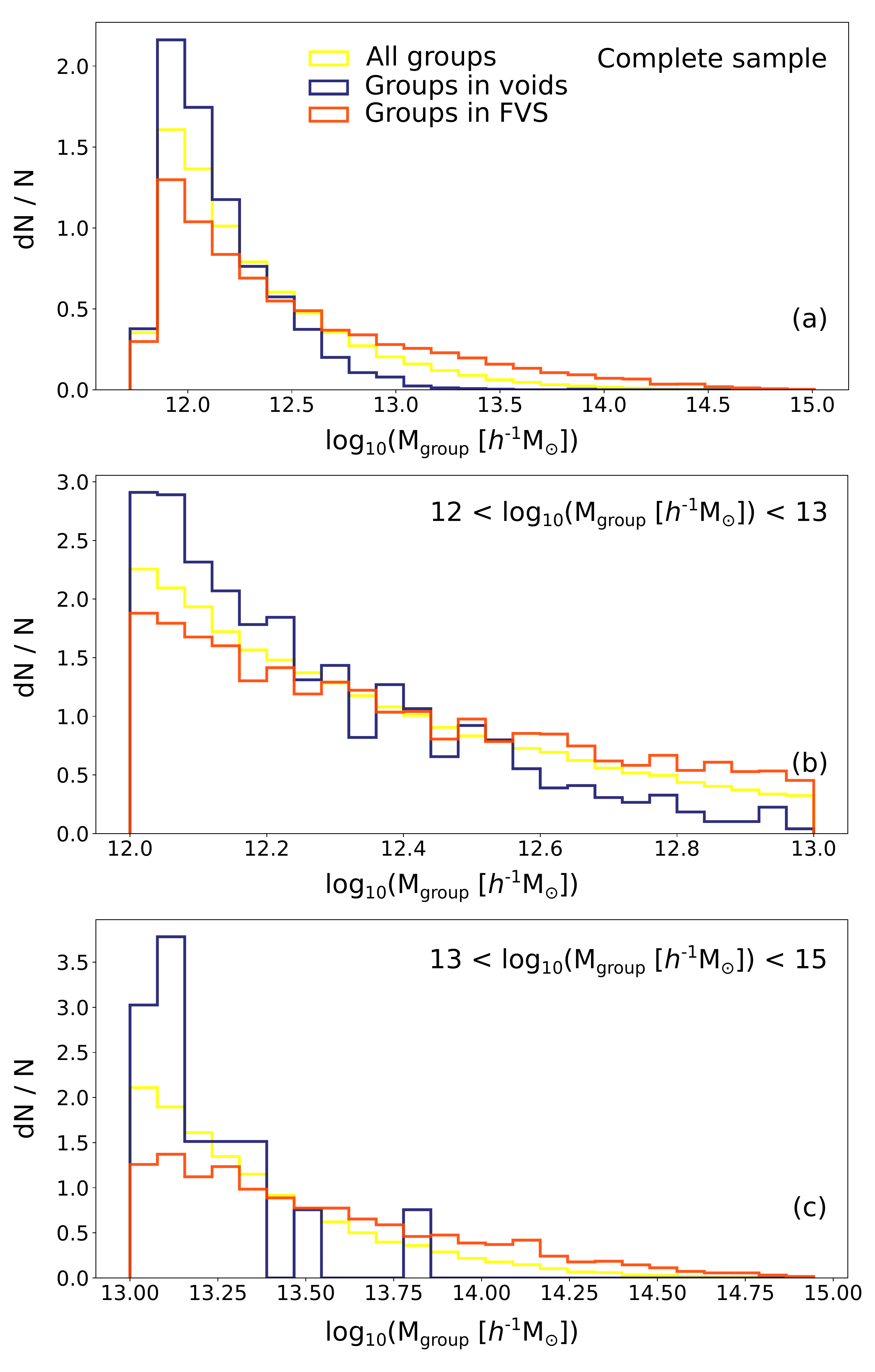}
\end{center}
\caption{\label{fig:hist_group} Normalized distribution of mass for the groups identified in the SDSS-DR12  by \citet{Rodriguez2020}. The complete sample is shown in panel (a), meanwhile in panels (b) and (c) are sub-samples with $12 < \log_{\rm 10}(M_{\rm group} [h^{-1}{\rm M}_{\odot}] < 13$ and $13 < \log_{\rm 10}(M_{\rm group} [h^{-1}{\rm M}_{\odot}] < 15$, respectively. In all cases the complete group samples (or sub-samples) are in yellow lines, the groups inside voids are in blue and the groups inside FVS are in red.}
\end{figure}

Concerning the dark matter component in these structures, we use the properties of the galaxy groups to estimate their features. We find that of the 98292 groups that set our main sample of galaxies, 1986 are in voids and 8233 in FVS.
The panel (a) of Fig. \ref{fig:hist_group} shows the normalised distribution of the estimated mass for the groups, $M_{\rm group}$, for the total sample of groups (yellow), the groups within voids (blue curve), and the groups within FVS (red curve).
Panels (b) and (c) show these same distributions for groups with $10^{12} \hmsun < M_{\rm group} < 10^{13} \hmsun$ and panel (c) for groups with $10^{13} \hmsun < M_{\rm group} < 10^{15} \hmsun$, respectively. 
For the latter subsample, voids only count with 22 galaxy groups and, consequently, their statistics is not so robust as in the other samples.
As expected, by inspection to these distributions, it is evident that voids have an excess of low-mass dark matter halos, while FVS show an excess of high-mass halos.

\section{HOD analysis in extreme density environments}
\label{sec:results}

To estimate the HOD we assume that each group represents a dark matter halo and compute the average number of galaxies in groups of a given mass, $\langle N_{\rm gal}~|~M_{\rm group}\rangle$.
Taking into account galaxy group membership it is straightforward to obtain the HOD by simply binning in group mass and calculate the average number of galaxies for each mass bin.

To study the behaviour of the HOD within the super-structures considered, we follow the procedure described above using only the groups that populate either voids or FVS.
Both super-structures were identified using galaxies as tracers, so both have incomplete groups in regions close to their boundaries.
In the case of voids, incomplete groups are removed from the sample before calculating the HOD, whereas for central galaxies residing within the volume of FVS, we consider their host groups.
Since their volumes are several times larger than that of the groups, this criterion does not considerably affect either the boundary conditions of any of the regions or the resulting HOD estimations.

To highlight the effects of these environments on the HOD, we define a third sample of galaxy groups that are not inside voids nor FVS.
We call this sample Field and we use this to repeat the analyses and measurements performed on the super-structure groups.

\begin{figure*}[h!]
\begin{center}
\includegraphics[width=\textwidth]{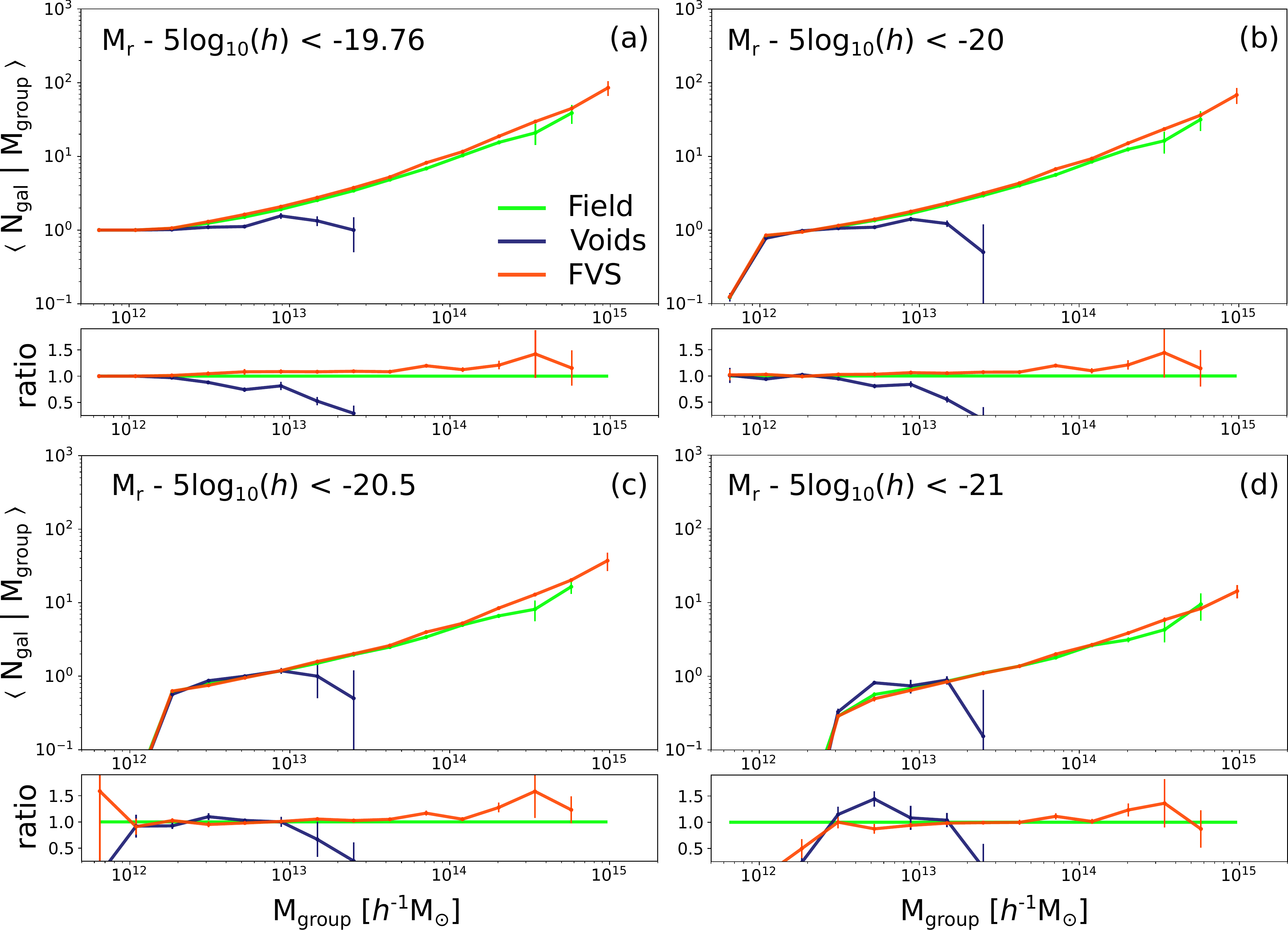}
\end{center}
\caption{\label{fig:HOD} HOD measured in four different magnitude ranges: $M_r - 5\log_{10}(h) < -19.76$, $-20$, $-20.5$ and $-21$ in panels (a), (b), (c) and (d), respectively. In green lines we show the HOD computed in the field, in blue the HOD inside voids and in red the HOD in FVS, as indicated in the key. We also show in the bottom of each panel the ratio between the HOD measured in the structures and the HOD in the field. All uncertainties were calculated by the standard jackknife procedure.}    
\end{figure*}

Fig. \ref{fig:HOD} shows the behaviour of the HOD within the voids, the FVS and in the Field, for different thresholds in absolute magnitude.
The absolute magnitude in the faintest $r$-band for which we can precisely estimate the HOD is $M_{\rm r} -5\log_{\rm 10}(h)=-19.76$, since from this point onwards the sample is no longer complete in volume and we lose the faint galaxies in the farthest groups.
Each panel shows at the top in green lines the HOD for the Field group sample, in red the HOD within FVS and HOD within voids in blue. 
Note that the last $M_{\rm group}$ bin is only populated by FVS galaxy groups. 
The statistical uncertainty for this point is low.
So this represents value information about the galaxy groups although there is not a counterpart in the Field or voids sample.  
The lower panels present the ratio between the measurements within these regions and the overall result.
Uncertainties in the calculations were computed with the jackknife technique. 
For this purpose, we separated the sample of halos into 50 equal sub-samples, and we computed HOD variations when we did not consider each one of these sub-samples in the measurements.
We also tested the results using 10, 100 and 150 sub-samples in the jackknife procedure, finding that, for 50 or more sub-samples, the variance values stabilise.
As it can be seen, for all the absolute magnitude ranges studied ($M_{\rm r} -5\log_{\rm 10}(h) = -19.76$, $-20$, $-20.5$ and $-21$) the HOD is systematically lower within the voids. In the FVS, on the other hand, the measurements are systematically higher. 
%

These results, even with the intrinsic differences of each sample, are completely consistent with those found in previous works on simulated data \citep{Alfaro2020,Alfaro2021}. 
Within the voids, the HODs decrease by up to 50$\%$ concerning the Field, while within the FVS they increase by up to 40$\%$.
Moreover, it is remarkable that within both regions systematic changes in the HOD are only observed from groups with masses larger than $\sim 10^{12} \hmsun$.
This is in agreement with the simulations, where the differences in the occupation of the dark matter halos in the simulations also only appear from halos with masses close to this critical value.
This is relevant because it seems to indicate that for masses below this critical value how the halo is populated does not depend on the large-scale environment.
However, above this mass, the environment starts to play an important role in the average number of galaxies in the halos. 

\subsection{Density dependence}
\label{sec:density}

Cosmic voids identification requires the setting of a threshold value for the integrated density contrast $\Delta_{\rm lim}$.
This parameter defines voids as regions with an integrated density contrast at $R_{\rm void}$ lower than the limiting value (see Sec. \ref{sec:voids_id}).
In this work we consider the usually adopted value $\Delta_{\rm lim} = -0.9$.
Given this rather strict restriction we expect an homogeneous behaviour across the cosmic void sample.
%
%
%

On the other hand, FVS are identified from a luminosity density field and unlike voids, no parameter determines the integrated density inside the FVS. For this reason, we may expect superstructures with different mean integrated densities.
Even within the same FVS, it is possible to find density variations between the outer and inner regions \citep{luparello_fvs_2011}.
In the identification procedure of FVS, (see \ref{sec:fvs_id}) we associate to each galaxy two parameters that characterise its local and global luminosity density: $\delta L_{\rm loc}$ and $\delta L_{\rm glob}$ corresponding to a $1\hmpc$ and a  $13\hmpc$ cube respectively.
By definition, all galaxies in FVS have $\delta L_{\rm loc} > 5.5$, however, there is a spread in their $\delta L_{\rm glob}$ values.
Galaxies with $\delta L_{\rm glob} < 5.5$ are likely located at the edge of FVS, while those with $\delta L_{\rm glob} > 5.5$ reside in the inner regions.
This parameter allows to study any dependence of the HOD on the density of the environment  by considering different regions within FVS.
%

\begin{figure*}[h!]
\begin{center}
\includegraphics[width=\textwidth]{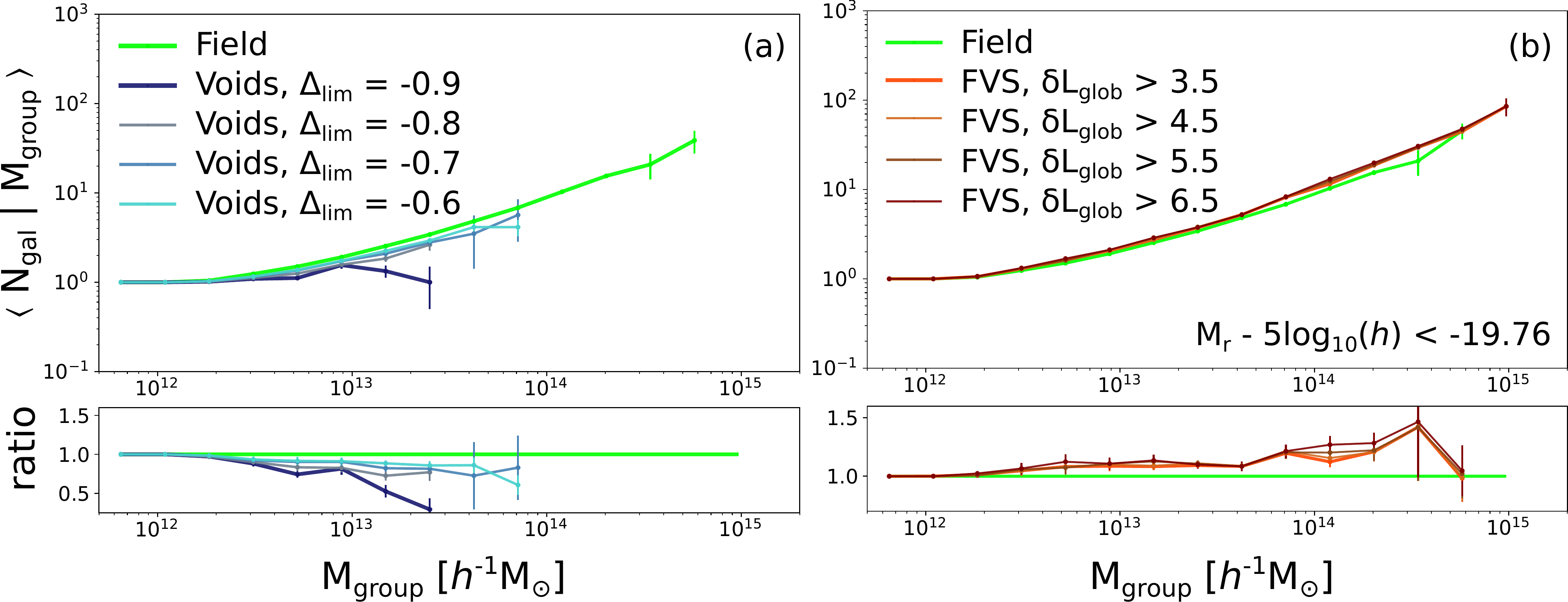}
\end{center}
\caption{\label{fig:HODdelta} Dependence of HOD with the identification parameters of voids and FVS for galaxies with $M_{\rm r} - 5\log_{\rm 10}(h) < -19.76$. In panel (a) we show the HOD measured in voids identified with different integrated density contrast threshold, from $\Delta_{\rm lim} = -0.9$ in dark blue to $\Delta_{\rm lim} = -0.6$ in light blue. In panel (b) we show the HOD measured in FVS with different global luminosity threshold, from $\delta L_{\rm glob} > 3.5$ in light red to $\delta L_{\rm glob} > 6.5$ in dark red. In both cases the HOD measured in the field is shown with the green lines. For a better quantification of the differences, inferior panels show the ratio between HOD measured in super-structures and the HOD of the field. All the uncertainties were calculated using the standard jackknife procedure.}
\end{figure*}

In panel (a) of Fig. \ref{fig:HODdelta} we show the dependence of the HOD with the $\Delta_{\rm lim}$ parameter of the voids. 
We identify voids with integrated density contrasts $\Delta_{\rm lim} = -0.6$, $-0.7$ and $-0.8$, and we measure within each sample the HOD and compare them with the result of the HOD measured in the field. 
The field HOD is shown with a green line, while the blue lines correspond to the HOD measured within the different voids catalogues.
In panel (b) of the same figure, we show the HOD within the FVS for different values of $\delta L_{\rm glob}$ in red lines. 
We also contrast these results with those of the field, showed with green lines.
In both cases, the panels below show the ratio between the different HODs and the HOD measured in the field.

As we can see, there is a clear dependence on the mean number of galaxies in the groups, the density limit value used to identify voids, and the regions with the highest luminosity density in the FVS. 
For both under-dense and over-dense regions, the differences in HOD increase as the density of the large-scale region surrounding the groups reaches extreme values. 
As we relax these conditions, the HOD becomes similar to the field.
We also performed both analysis for $M_{\rm r} - 5\log_{\rm 10}(h) = -20$, $-20.5$, and $-21$, finding similar results.

\subsection{Dependence on structure properties}
\label{sec:volume}

We have previously analysed the correlation between the average number of galaxies in groups and the large-scale structure. 
In this subsection we search for possible dependence of the HOD on intrinsic properties of voids and FVS.

\begin{figure}[h!]
\begin{center}
\includegraphics[width=\columnwidth]{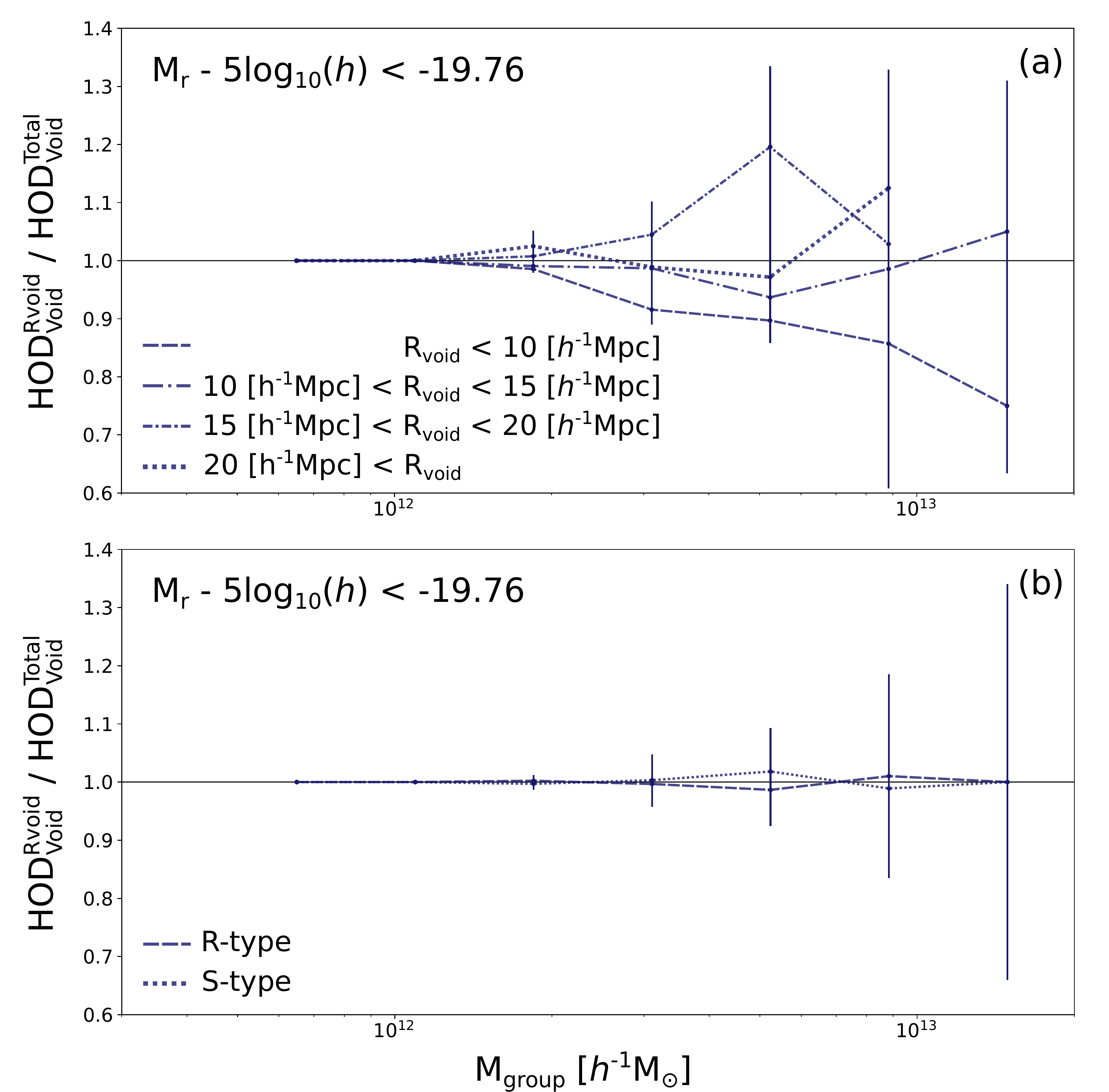}
\end{center}
\caption{\label{fig:Rvoid} {\it Panel (a)}: Dependence of the HOD measured inside cosmic voids with void sizes $R_{\rm void}$. Different line types correspond to different void radius intervals as indicated in the key. {\it Panel (b):} Same as panel (a) but considering the void dynamical classification, where S-type voids are shown with dotted lines and R-type voids with dashed lines. For both panels, all the uncertainties were calculated by the standard jackknife procedure.}
\end{figure}

Besides the criterion of a threshold galaxy density for voids, these can be characterised by their size and the density of the surrounding environment.
According to the latter criterion, voids can be classified as R-type and S-type \citep{ceccarelli_clues_2013, paz_clues_2013}. 
R-type voids are surrounded by large-scale under-dense regions, while S-types are embedded in global over-dense regions.
To further explore the behaviour of the HOD according to these properties, we divide our group sample in voids according to the void radius and void-type classification. Then, we measure the mean number of galaxies per group in the same way as described in Sec. \ref{sec:results}.
The relative HOD results for galaxies with $M_{\rm r} - 5\log_{\rm 10}(h)<-19.76$ can be seen in Fig. \ref{fig:Rvoid}. 
The panel (a) shows the ratio of the HOD for four subsamples of voids according to their radius values : $R_{\rm void}<10\hmpc$, $10\hmpc<R_{\rm void}<15\hmpc$, $15\hmpc<R_{\rm void}<20\hmpc$ and $20\hmpc<R_{\rm void}$, with respect to the HOD of the full void catalogue.
Panel (b), on the other hand, shows the ratios of the HOD in R-type (dashed line) and S-type (dotted line) voids concerning the HOD of the full catalogue of groups within voids.
For both cases, we find is no clear dependence of the HOD behaviour on the size nor type of the void.
Thus, the HOD has a reasonable universal behavior in voids.

\begin{figure}[h!]
\begin{center}
\includegraphics[width=\columnwidth]{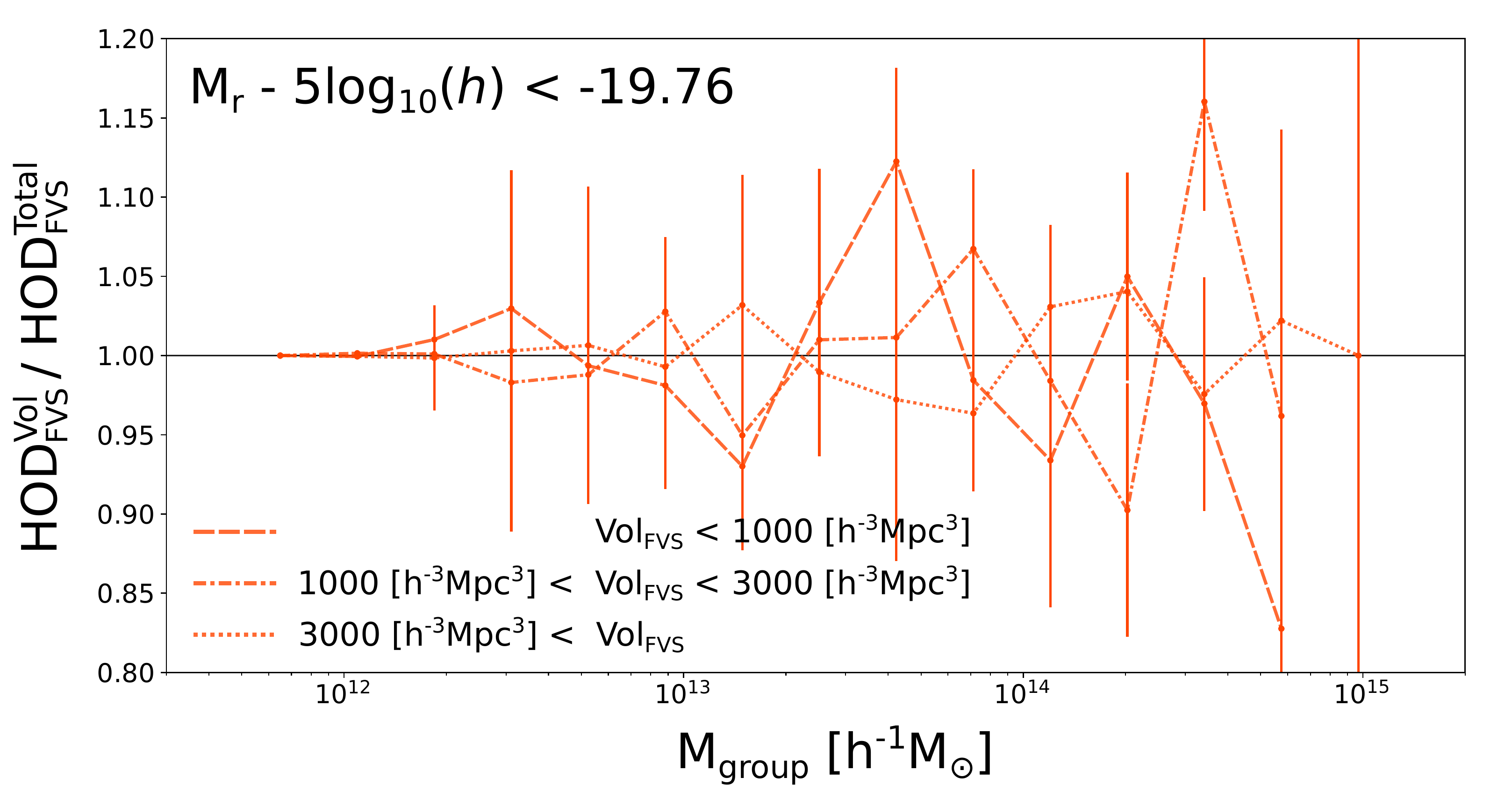}
\end{center}
\caption{\label{fig:HOD_Vol} Ratios between the HOD for different FVS volume ranges (indicated in the key figure) and the complete FVS sample. All the uncertainties were calculated by the standard jackknife procedure.}
\end{figure}

For FVS, we have studied a possible dependence on the volume of these super-structures.
For this aim, we divide the sample into three bins: $V_{\rm FVS}< 2500~h^{-3}{\rm Mpc}^3$, $2500~h^{-3}{\rm Mpc}^3 < V_{\text{FVS}} < 5000 ~h^{-3}{\rm Mpc}^3$ , and $5000~h^{-3}{\rm Mpc}^3 < V_{\text{FVS}}$, and we compute the HOD in each of them. 
The results are given in Fig. \ref{fig:HOD_Vol}, where the different lines correspond to the ratio between the HOD of each bin in FVS volume and the HOD measured for the full FVS sample.
Again here, there is not a neat evidence of a dependence between the volume of the super-structures and the HOD behaviour.

Both voids and FVS observational results are in agreement with our predictions in simulated data \citep{Alfaro2020,Alfaro2021}, where no evidence of a correlation between HOD variations and intrinsic properties of large-scale regions is detected.
%

\section{Halo formation time groups} 
\label{sec:stellar}

In \cite{Alfaro2020} and \cite{Alfaro2021} we found evidence that simulated dark matter halos have different formation times when they reside in voids or in FVS. 
In addition, there are correlations between the density of the large-scale environment surrounding halos, their formation times, and the HOD.
Halos in voids are younger and have lower HODs than average, while halos in FVS are older and have higher HODs.

In the simulations, it is possible to follow the formation history of each dark matter halo to determine its formation time.
In this observational date set we will use as an age indicator the onset time of star formation in the central galaxy of the group.
This parameter is predicted by the stellar mass model of \cite{Maraston2006}, and we call it $T_{\star}$.
We define the central galaxy of each group as the brightest object and calculate the cumulative distribution of $T_{\star}$ for the Field sample and the group samples in voids and FVS.

\begin{figure}[h!]
\begin{center}
\includegraphics[width=\columnwidth]{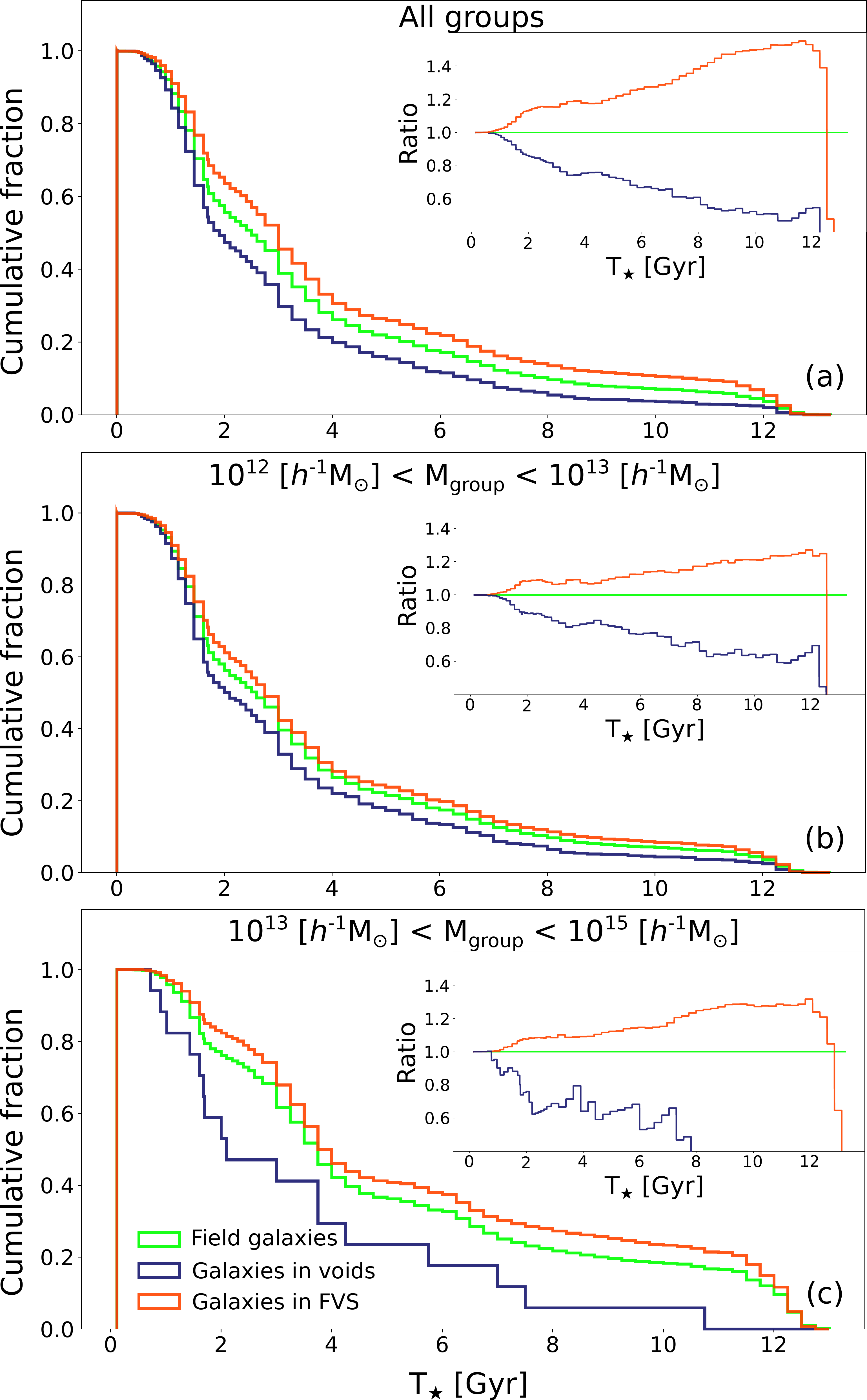}
\end{center}
\caption{\label{fig:HistTstar} {\it Top}. Cumulative distribution of $T_\star$ parameter for Field galaxies (green), void galaxies (blue) and FVS galaxies (red). Panel (a) show the distributions for the complete galaxy sample, meanwhile panels (b) and (c) are for the galaxies residing in groups with $10^{12} \hmsun < M_{\rm group} < 10^{13} \hmsun$ and $10^{12} \hmsun < M_{\rm group} < 10^{13}\hmsun$, respectively. Inset panels show the ratio between the cumulative fraction of $T_\star$ for galaxies inside voids/FVS and Field galaxies. }
\end{figure}

Panel (a) in Fig. \ref{fig:HistTstar} shows the cumulative distribution of $T_{\star}$ parameter for the group samples in the field (green), voids (blue) and FVS (red).
The inset panel in this figure presents the ratio between super-structures and Field.
Using the same format, panels (b) and (c) show the same but for groups with $10^{12}\hmsun<M_{\rm group}<10^{13}\hmsun$ and $10^{13}\hmsun<M_{\rm group}<10^{14}\hmsun$, respectively.

In all group samples, it is clear that objects within voids have $T_{\star}$ lower than the mean.
On the contrary, clusters residing in FVS have higher $T_{\star}$ values.
Generally speaking we may say that star formation started earlier in FVS than the average contrary to voids which show a more recent onset of the star formation process. 
These results are in agreement with the differences in the formation times of dark matter halos found in the synthetic data.
Thus, we confirm from the observational side, evidence for a correlation between the HOD, the halo large-scale density environment and the formation time of the halos.

\subsection{HOD in function of $T_{\star}$} 
\label{sec:HODstellar}

\begin{figure*}[h!]
\begin{center}
\includegraphics[width=\textwidth]{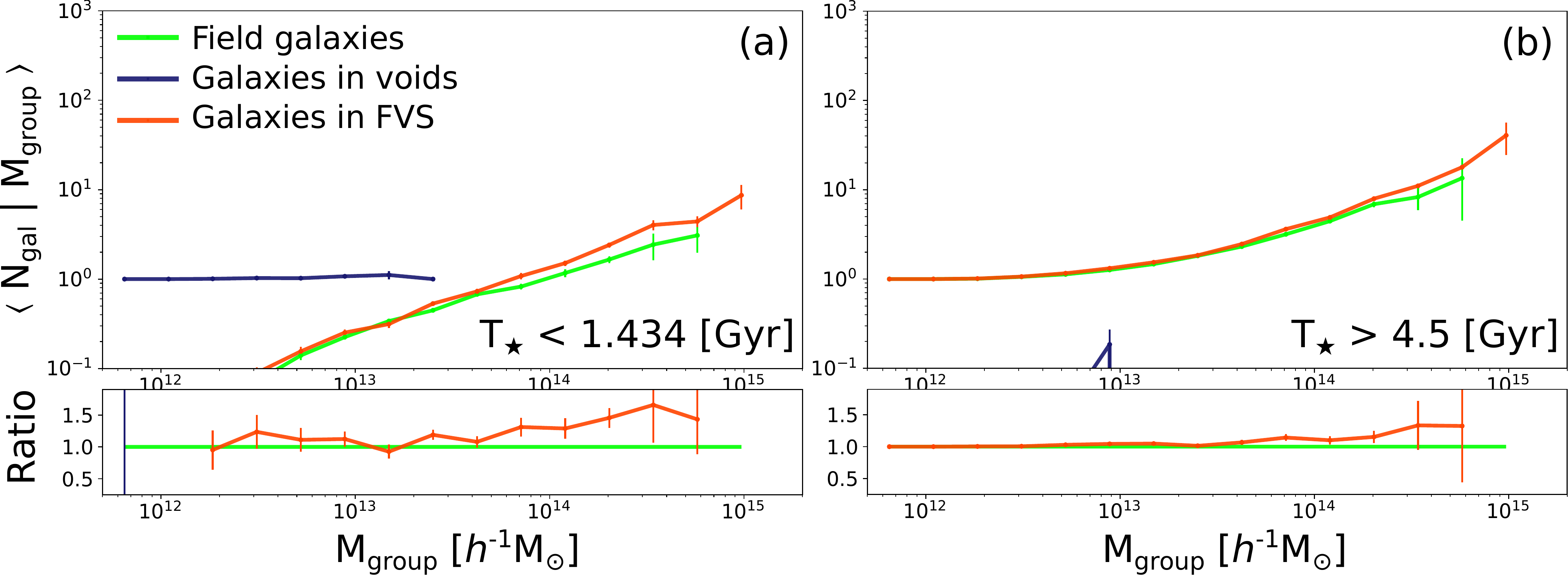}
\end{center}
\caption{\label{fig:HOD_Tstar} HOD for galaxies in the first and fourth quartile of $T_\star$, panels (a) and (b) respectively. In both cases, Field galaxies are in green lines, void galaxies in blue and FVS galaxies in red. All uncertainties were calculated with jackknife procedure.}
\end{figure*}

We have already seen that the galaxy groups populating voids and FVS have different average number of galaxies as well as different star formation times of their central galaxy.
In this section, we further explore the relation betweem these two properties and calculate the HOD as a function of the $T_{\star}$ parameter, rather than their r-band luminosty.
%

%
For this purpose, we sort the entire sample of galaxies (including central and satellite galaxies) by their $T_{\star}$ and consider separately those in the first and fourth quartiles.
This is equivalent to take the 25$\%$ fraction of the oldest/youngest galaxies which correspond to those with $T_{\star} \leq 1.434 {\rm Gyr}$ and $2.75 {\rm Gyr} \leq T_{\star}$ respectively.
In order to analyse these two galaxy populations and the way they are affected by environment, we calculate their HOD within cosmic voids, FVS and the Field.
The procedure is the same as described in Sec. \ref{sec:results} except that here we only consider those galaxies belonging to these two quartile subsamples.

Panel (a) of Fig. \ref{fig:HOD_Tstar} shows the measured HOD for the $25\%$ younger galaxies, the blue line represents the result within the voids, the red line within the FVS and the green line in the Field.
We can see that in the case of the FVS the environment affects in the same way as described in Sec \ref{sec:results} and the number of young galaxies per group increases by almost $50\%$ concerning the Field.
In addition, we see that low-mass groups only host young galaxies when they reside inside cosmic voids.
%
%
In Panel (b) we show the HOD results for the $25\%$ oldest galaxies in the sample. 
We see here that groups in cosmic voids lack this old population , while on the contrary, FVS groups show a higher number of old satellites as compared to Field groups.

In general, we can conclude that for groups residing in FVS the HOD variation with respect to the Field is similar for young and old galaxies.
We have confirmed this conclusion varying the definitions of young and old galaxies with different $T_{\star}$ thresholds finding results totally consistent with those described above.

\section{Summary and conclusions}
\label{sec:conclusions}

The HOD is a powerful tool linking galaxies to their host dark matter halos.
In this work, we use observational data to study HOD behaviour in different large-scale environments with extreme density values.
We have considered cosmic voids and FVS as low/high density super-structures.
We use a volume complete sample of the SDSS-DR12, restricted to $M_{\rm r} - 5\log_{\rm 10}(h) < -19.76$ galaxies, and the galaxy group catalogue of \cite{Rodriguez2020}.
We have applied voids and FVS identification algorithms to define three galaxy groups samples: Groups in Voids,  FVS and Field (groups that are not located in any of both regions).

We find a statistically significant difference between the HOD of groups residing in these two environments.
Inside cosmic voids, the HOD is consistent with a decrease of $\sim 50\%$ in the mean number of satellites with respect to the field HOD.
Conversely, for groups in FVS, the HOD shows an increase up to $\sim 40\%$. 
Also, we note that the FVS sample is the one that contains the groups with the highest mass.
These results are present for all luminosity ranges explored.

For both types of environments, we find a clear dependence of HOD on galaxy density. 
In cosmic voids, the HOD difference with respect to the field increase as the value of the $ \Delta_{\rm lim} $ parameter lowers, ie. towards more empty voids.
On the other hand, in FVS the differences are larger in the central, densest regions with highest values of the $\delta L_{\rm glob}$ parameter. 

The aforementioned results are present only for massive groups, ie. with masses greater than $\sim 10^{12}\hmsun$. 
Irrespective of the large--scale environment, the HOD for groups with masses lower than $\sim 10^{12}\hmsun$ exhibit no variations.
This indicates that for these groups, the formation of galaxies is nearly independent of the large-scale environment density.
In Sec.\ref{sec:est_prop}, we also find no evidence that the HOD variations depend on the intrinsic properties of the super-structures.
Inside the voids, the HOD is independent of the radii and the surrounding structure.
For the FVS, the HOD shows no dependence with the superstructure volume.
All this results are consistent with the observed in simulated data in \cite{Alfaro2020} and \cite{Alfaro2021}. 

Finally, in Sec. \ref{sec:stellar} we show that the central galaxy of groups within voids has an onset time of star formation ($T_{\star}$ parameter) lower than their counterpart in Field groups.
In FVS, the central galaxies of groups show systematically higher star formation times.
These results could be related to the differences in the assembly time of dark matter halos, as reported in previous works.
In simulations, halos inside FVS formed earlier than average, opposite to the more recent assembly inside  cosmic voids. 

For a more detailed analysis, we further explore the HOD for the $25\%$ fraction of oldest and youngest galaxies respectively, based on the galaxy $T_{\star}$ values.
We find that the youngest galaxies within low-mass groups ($< 10^{12} \hmsun$) are limited to the inner regions of cosmic voids. 
The youngest galaxies within high-mass groups ($> 10^{12} \hmsun$) are mainly found in the Field and in FVS. 
Thus, there is a connection between astrophysical galaxy properties and the HOD regarding environment.

Regardless group mass, the oldest galaxies mainly reside in FVS and in the  Field.
Thus, cosmic void galaxies lack old stellar populations irrespective of their local environment.
Galaxies with evolved stellar population are mainly located in FVS and Field groups with a diverse range of the mass.
Our work provides evidence of large--scale environment combined effects manifest both in HOD as well as galaxy astrophysics.
Similar studies in future deeper surveys may highlight the interplay between HOD and galaxy astrophysical properties at early epochs where density contrasts associated to FVS and cosmic voids are lower than at the present.

\begin{acknowledgements}
 This work was partially supported by Agencia Nacional de Promoción Científica y Tecnológica (PICT 2015-3098, PICT 2016-1975), the Consejo Nacional de Investigaciones Científicas y Técnicas (CONICET, Argentina) and the Secretaría de Ciencia y Tecnología de la Universidad Nacional de Córdoba (SeCyT-UNC, Argentina).

 Funding for the SDSS and SDSS-II has been provided by the Alfred P. Sloan Foundation, the Participating Institutions, the National Science Foundation, the U.S. Department of Energy,
the National Aeronautics and Space Administration, the Japanese Monbukagakusho, the Max Planck Society, and the Higher Education Funding Council for England. The SDSS Web Site is http://www.sdss.org/. The SDSS is managed by the Astrophysical Research Consortium for the Participating Institutions. The Participating Institutions are the American Museum of Natural History, Astrophysical Institute Potsdam, University of Basel, University of Cambridge, Case Western Reserve University, University of Chicago, Drexel University, Fermilab, the Institute for Advanced Study, the Japan Participation Group, Johns Hopkins University, the Joint Institute for Nuclear Astrophysics, the Kavli Institute for Particle Astrophysics and Cosmology, the Korean Scientist Group, the Chinese Academy of Sciences (LAMOST), Los Alamos National Laboratory, the Max-Planck-Institute for Astronomy (MPIA), the Max-Planck-Institute for Astrophysics (MPA), New Mexico State University, Ohio State University, University of Pittsburgh, University of Portsmouth, Princeton University, the United States Naval Observatory, and the University of Washington.

\end{acknowledgements}

\bibliographystyle{aa}
\bibliography{references}

\end{document}